\documentstyle[aps,prl]{revtex}
\def\update{8 March 2001}
\def\version{[PRL]}

\tighten

\def\ja{{\hat{\jmath}}}

\begin{document}
\preprint{}
\draft

\wideabs{

\title{Reply to ``Comment about pion electroproduction and the axial form factors"}

\author{H. Haberzettl}

\address{Center for Nuclear Studies, Department of Physics,
         The George Washington University, Washington, D.C. 20052}

\date{\update}

\maketitle

\begin{abstract}
It is shown that comments by Guichon \cite{guichon},
and also by Bernard, Kaiser, and Mei{\ss}ner \cite{bkm01},
regarding my recent criticism \cite{hh00ax}
of how the axial form factor is supposed to enter pion electroproduction
do not address the main point of my argument and therefore are irrelevant.
\end{abstract}
\pacs{PACS numbers: 13.60.Le \hfill {\tiny \version}}
}

In a comment, Guichon \cite{guichon} criticizes my recent paper \cite{hh00ax}
which shows that one cannot measure the axial form factor in pion production measurements at threshold.
The same criticism, albeit in a less civilized manner, but based on essentially the same argument,
was also raised by  Bernard, Kaiser, and Mei{\ss}ner \cite{bkm01}.
I will show here that none of these criticisms does address the main point of my argument.
I still maintain therefore, that the conclusions of \cite{hh00ax} stand as stated.

Both arguments \cite{guichon,bkm01}  concern the splitting of the axial current $\ja^\mu_A$
into
$\ja^\mu_A=\ja^\mu_{A,W}+\ja^\mu_{A,H}$,
where \cite{hh00ax}
\begin{mathletters}\label{jaxsplit}
\begin{eqnarray}
\ja^\mu_{A,W} &=& -\gamma_5\left[ \gamma^\mu
                   +(p'-p)^\mu\frac{2m}{t}  \right] G_A  \frac{\tau}{2}
            \;,\label{jaxw}
\\
\ja^\mu_{A,H} &=&  -f_\pi (p'-p)^\mu \frac{\mu^2}{t} \frac{1}{t-\mu^2}
                    \gamma_5 G_t \tau \;\label{jaxh}
\end{eqnarray}
\end{mathletters}
separate the dependence on the axial form factor $G_A$ and the $\pi NN$ form factor
$G_t$.
The $\pi NN$ form factor enters the expression for the axial current upon invoking
the PCAC constraint and eliminating the pseudoscalar form factor $G_P$.
The criticisms center now on the occurrence of the $1/t$
singularities in both of the terms of (\ref{jaxsplit}).

Clearly, this splitting, referred to as ``strange" in Ref. \cite{guichon}, is
without question algebraically valid, even if it creates seemingly
artificial poles 1/t in both terms. Written differently,
\begin{eqnarray}
\ja^\mu_A&=& -\gamma_5 \gamma^\mu G_A  \frac{\tau}{2}
           -f_\pi  \frac{(p'-p)^\mu}{t-\mu^2} \gamma_5 G_t \tau
\nonumber\\
&& { }+ (p'-p)^\mu \gamma_5\frac{f_\pi G_t - m G_A}{t} \tau
\;,\label{jaxlimit}
\end{eqnarray}
these singularities for $t=0$ are recast in the well-defined $\frac{0}{0}$ situation contained
in the last term in view of the validity of the Goldberger--Treiman
relation.

This splitting clearly shows that in the chiral limit of vanishing pion mass, the
hadronic part of the axial current, Eq.\ (\ref{jaxh}), vanishes. The $1/t$
pole of the remaining weak part, Eq.\ (\ref{jaxw}), now cannot be cancelled
any longer and thus becomes physically relevant in the chiral limit.
Furthermore, this splitting allows for a most detailed description
of how the electromagnetic (vector) current couples to the axial
current, as given in Eq. (14) and as depicted in Fig. 3, both of Ref.\ \cite{hh00ax}. I presume
that this is what is referred to as the ``trap in the
reasoning" \cite{guichon}. There is, however, nothing wrong with this procedure.
As mentioned in my paper, it can be shown quite directly by
employing Green's function techniques that since the splitting
$\ja_A=\ja_{A,W}+\ja_{A,H}$ is valid, the coupling to the electromagnetic
current does indeed produce Eq. (14). For this result to obtain, it is irrelevant
whether one uses Eq.\ (\ref{jaxsplit}) here or the nonsingular form (\ref{jaxlimit}).
The advantage of the former, in my opinion, is simply that it allows for a more
straightforward graphical interpretation (see Fig.\ 3 of \cite{hh00ax}).
The subsequent steps in \cite{hh00ax} leading
to Eq.\ (19) and finally to Eq.\ (23) are purely algebraic. In
particular, they show that the complete cancellation of the $G_A$
dependence obtains {\it before} the $q=0$ limit is taken. The fact that $\mathcal{W}$
[of Eq. (20)] vanishes even if the nucleon has structure (which I
don't dispute and which I didn't discuss in my Letter for lack of
space) is irrelevant in this context. What is relevant is the
statement after Eq.\ (23) that ``on the right-hand side the
dependence on $G_A$ cancels completely even before the limit is
taken". This crucial statement can very easily be verified by
direct evaluation of the expression in the square brackets on the
right-hand side of Eq.\ (23) along the lines given in my Letter. If
the author of Ref.\ \cite{guichon} believes that this evaluation is in error, he should be
able to point out precisely which of the steps outlined in detail
in my paper is in error. The splitting discussed above clearly is not
at issue in this respect.

The arguments of Ref.\ \cite{bkm01} are based on arbitrarily demanding that the factor
$\mu^2/t$ in Eq.\ (\ref{jaxh}) above should be replaced by unity. This
would essentially
drop part of the last term in Eq.\ (\ref{jaxlimit})  containing the $\frac{0}{0}$
situation and would indeed lead to the result desired by these authors.
There is, however, nothing which justifies this arbitrary replacement
and this argument evidently is in error. The conclusions of my paper,
therefore, stand as stated.

\end{document}